\documentclass[aps,twocolumn,amsmath,amssymb,preprintnumbers]{revtex4}
\usepackage{amsmath} \usepackage{amsfonts} \usepackage{amssymb}
\usepackage{bbm}
\usepackage{epsfig}
\usepackage{graphics}
\usepackage{graphicx}
\newcommand{\be}{\begin{equation}}
\newcommand{\ee}{\end{equation}}
\newcommand{\ba}{\begin{eqnarray}}
\newcommand{\ea}{\end{eqnarray}}
\newcommand{\nn}{\nonumber}

\newcommand{\iD}{^{(D)}}

\newcommand{\A}{A^{(k)}}
\newcommand{\kl}{\langle}
\newcommand{\kr}{\rangle}
\newcommand{\C}{^{(C)}}
\newcommand{\Ai}{^{(A)}}
\newcommand{\B}{^{(B)}}
\newcommand{\Q}{^{(Q)}}
\newcommand{\tr}{\textup{tr}}

\begin{document}

\title[ ]{Emergence of quantum mechanics from classical statistics}

\author{C. Wetterich}
\affiliation{Institut  f\"ur Theoretische Physik\\
Universit\"at Heidelberg\\
Philosophenweg 16, D-69120 Heidelberg}

\begin{abstract}
The conceptual setting of quantum mechanics is subject to an ongoing debate from its beginnings until now. The consequences of the apparent differences between quantum statistics and classical statistics range from the philosophical interpretations to practical issues as quantum computing. In this note we demonstrate how quantum mechanics can emerge from classical statistical systems. We discuss conditions and circumstances for this to happen. Quantum systems describe isolated subsystems of classical statistical systems with infinitely many states. While infinitely many classical observables ``measure'' properties of the subsystem and its environment, the state of the subsystem can be characterized by the expectation values of only a few probabilistic observables. They define a density matrix, and all the usual laws of quantum mechanics follow. No concepts beyond classical statistics are needed for quantum physics - the differences are only apparent and result from the particularities of those classical statistical systems which admit a quantum mechanical description. In particular, we show how the non-commuting properties of quantum operators are associated to the use of conditional probabilities within the classical system, and how a unitary time evolution reflects the isolation of the subsystem.
\end{abstract}

\maketitle
A realization of quantum mechanics as a classical statistical system may shed new light on the conceptual interpretation of experiments based on entanglement, as teleportation or quantum cryptography \cite{Ze}. One may even speculate that steps in quantum computing \cite{Zo} could be realized by exploiting classical statistics. Recently, classical statistical ensembles that are equivalent to four-state and two-state quantum mechanics have been constructed explicitely \cite{CW1,CW2}.  This constitutes a proof of equivalence of few-state quantum statistics and classical statistics for infinitely many degrees of freedom. In view of the particular manifolds of classical states for these examples one may wonder, however, if quantum statistical systems are very special cases of classical statistics, or if they arise rather genuinely under certain conditions. In this note we argue that quantum statistics can indeed emerge rather generally if one describes small ``isolated'' subsystems of classical ensembles with an infinity of states.

An atom in quantum mechanics is an isolated system with a few degrees of freedom. This contrasts with quantum field theory, where an atom is described as a particular excitation of the vacuum. The vacuum in quantum field theory is a complicated object, involving infinitely many degrees of freedom. In a fundamental theory of particle physics, which underlies the description of atoms, collective effects as spontaneous symmetry breaking are crucial for its understanding. Our treatment of atoms in the context of classical statistics is similar to the conceptual setting of quantum field theory. A classical statistical system with infinitely many states describes the atom and its environment or the atom and the vacuum. The quantum statistical features become apparent if one concentrates on a subsystem that describes the isolated atom. 

Our classical statistical description of quantum systems has three crucial ingredients. (1) {\em Probabilistic observables} have in a given quantum state only a probabilistic distribution of possible measurement values, rather than a fixed value as for classical observables in a given state of the classical ensemble. Probabilistic observables obtain from classical observables by ``integrating out'' the environment degrees of freedom. The fact that this map is not invertible avoids conflicts with the Kochen-Specker theorem \cite{KS,Str}.  The probabilistic nature of the observables can be understood as a result of ``coarse graining of the information'', starting from deterministic observables on a suitable level. Alternatively, this concept may be used as a basic definition of observables \cite{POB}.

(2) {\em Conditional probabilities} are used for a computation of the probabilities for the possible outcomes of two measurements of observables $A$ and $B$. In particular, if $B$ is measured after $A$, the outcome of the measurement of $B$ depends on the previous measurement of $A$. Conditional probabilities induce the concept of {\em quantum correlations}, and we propose that quantum correlations rather than classical correlations should be used for a description of two measurements. This avoids conflicts with Bell's theorem \cite{Bell,Z}. The classical correlations are actually not uniquely defined for the quantum system - they depend on properties of the environment. 

(3) {\em The unitary time evolution} of quantum mechanics is a special case of a more general classical evolution which can also describe the phenomena of decoherence and syncoherence. We propose that the unitary time evolution reflects the isolation of the subsystem and corresponds to a partial fixed point (or better ``fixed manifold'') of the more general classical time evolution.

A classical statistical description of quantum mechanics should not be confounded with a deterministic description. The no go theorems for large classes of local deterministic ``hidden variable theories'' remain valid. We rather take the attitude that a probabilistic approach is appropriate for the basic setting of both the quantum and the classical world. Deterministic behavior is a particular case (albeit rather genuine) which originates from the collective properties of many degrees of freedom, as the laws of thermodynamics, or the motion of planets which are composed of many atoms. 

\medskip
\noindent
{\bf Probabilistic observables}

The most crucial effect of the embedding of the subsystem into classical statistics for infinitely many states is the appearance of probabilistic observables for the description of the subsystem. For a given state of the subsystem - which will be associated with a quantum state - they have a probability distribution of values rather than a fixed value as for the standard classical observables.  For subsystems that are equivalent to $M$-state quantum mechanics the spectrum of the possible outcomes of measurements for the probabilistic observables contains at most $M$ different real values $\gamma_a$. In a given quantum state the probabilistic observable is characterized by probabilities $w_a$ to find $\gamma_a$, where $w_a$ depends on the state. The simplest example are two-level observables, which can resolve only one bit, such that $\gamma_1=1~,~\gamma_2=-1$. 

Let us consider a subsystem that can be described by $n$ discrete classical two-level-observables $A^{(k)}$. The subsystem is embedded into a classical statistical ensemble with infinitely many states labeled by $\tau$. On this level the observables $A^{(k)}$ are standard classical or deterministic observables. They can only take fixed values $A^{(k)}_\tau=\pm 1$ for any state $\tau$ of the classical statistical ensemble. We assume that these observables form a basis in a sense to be specified later. The simplest quantum mechanical analogue for $n=3$ is two-state quantum mechanics, with $A^{(k)}$ corresponding to three orthogonal ``spins'' in an appropriate normalization. This may be viewed as an atom with spin one half where only the spin degree of freedom is resolved, as for example in Stern-Gerlach type experiments. For arbitrary $n$, we denote the expectation value or average of $A^{(k)}$ by $\rho_k$, 
\be\label{1}
\rho_k=\langle A^{(k)}\rangle=\sum_\tau p_\tau A^{(k)}_\tau~,~-1\leq \rho_k\leq 1.
\ee

Here we have formulated $\A$ as classical observables with classical probabilities 
$p_\tau\geq 0$ for the states $\tau$ of the classical statistical ensemble. As usual, one has $\sum_\tau p_\tau=1$. 
(Not all classical probability distributions $\{p_\tau\}$ correspond to quantum systems, and we will discuss restrictions below.) The probabilistic observables associated to these classical observables have probabilities $w^{(k)}_\pm=(1\pm\rho_k)/2$ to find $\gamma_{1,2}=\pm 1$. We can compute $w^{(k)}_+$ by summing the probabilities $p_\tau$ over all classical states for which $\A_\tau=1$, and similar for $w^{(k)}_-$. This maps the classical observable, as characterized by the values $\A_\tau$ in every classical state $\tau$, to a probabilistic observable characterized by $\gamma^{(k)}_a$ and $w^{(k)}_a$. 

The map from the classical observables to the probabilistic observables is not invertible. A different classical observable $A'^{(k)}$, with different values $A'^{(k)}_\tau$ in the classical states, may be mapped to the same probabilistic observable $\A$. For this it is sufficient that the probabilities $w^{(k)}_\pm$ are the same for every probability distribution $\{p_\tau\}$ which corresponds to a quantum system. This lack of invertibility avoids conflicts of our classical statistical description with the Kochen-Specker theorem \cite{KS}, as we we will discuss at the end of this note. Since the classical observables $\A_\tau$ contain much more information than the associated probabilistic observables, they ``measure'' properties of both the quantum system and its environment. The transition to probabilistic observables ``integrates out'' the environment degrees of freedom \cite{CW2}.

\medskip\noindent
{\bf Quantum system and system observables}

We will assume that the $n$ numbers $\rho_k$ are the only information that is needed for a computation of expectation values for the ``system observables'' of the subsystem. In this sense, the state of the subsystem is characterized by the $n$ expectation values of the basis observables $A^{(k)}$. Only a very limited amount of the information contained in the probability distribution $\{p_\tau\}$ for the total system is needed for the subsystem. Quantum mechanics is ``incomplete statistics'' in the sense of ref. \cite{3}. More precisely, system observables are those classical observables that lead to probabilistic observables $A$ for which the probabilities $w_a$ can be computed in terms of $\{\rho_k\}$. Then the expectation values of functions $f(A)$ can also be computed from $\{\rho_k\}$, $\kl f(A)\kr=\sum_a f(\gamma_a)w_a(\rho_k)$. We will assume that the relation between $w_a$ and $\rho_k$ is linear. 

Our first question concerns a classification of possible system observables for the subsystem. It is straightforward to define rescaled observables $c A^{(k)}$ by $(c A^{(k)})_\tau=c A^{(k)}_\tau,\langle c A^{(k)}\rangle=c\rho_k$.  Furthermore, we can trivially shift the observable by a piece $e_0$ proportional to the unit observable. The rescaled and shifted observables $A$ obey
\be\label{2}
\langle A\rangle=\rho_ke^{(A)}_k+e^{(A)}_0,
\ee
where repeated indices are summed. Here we associate to each $A^{(k)}$ an $n$-dimensional unit vector $e^{(k)}$ with components $e^{(k)}_m=\delta^k_m$.  Then the vector $e^{(A)}$ reads $e^{(A)}=c e^{(k)}$ if $A=c A^{(k)}$. One may use $c=\hbar/2$ if $A$ describes a spin with standard units of angular momentum. Other units may be employed for alternative interpretations, as for example occupation number $n=(1+A^{(3)})/2$ which equals one for occupied and zero for empty. (Contrary to widespread belief, $\hbar$ is not a genuine property of quantum mechanics, but rather an issue of units.) In the following we mainly discuss observables with $e^{(A)}_0=0$.

For an arbitrary system observable $A$ we may write the expectation value $\kl A\kr=\sum_a\gamma_a w_a(\rho_k)$, with $w_a$ depending linearly on $\rho_k$, as a linear combination $\langle A\rangle=\sum_kc_k\langle A^{(k)}\rangle$. (If necessary, we substract an appropriate constant shift.) This should hold for all probability distributions $\{p_\tau\}$ which describe the quantum system. (Such observables may correspond to rotated spins in the quantum mechanical analogue.) We can associate to each such observable the vector $e^{(A)}=\sum_kc_k e^{(k)}~,~e^{(A)}_k=c_k$, such that eq. \eqref{2} remains valid. Explicit constructions of such probabilistic observables can be found in \cite{CW2}. The possibility to write the expectation value in the form (2) is a necessary, albeit not sufficient condition for a classical observable to be a system observable. 

Probabilistic observables are characterized by the spectrum $\{\gamma_a\}$ of possible measurement values, and the associated probabilities $w_a$. The multiplication of the observable by a constant $c$ and the addition of a piece proportional to the unit observable are always defined by $\gamma_a\to c\gamma_a+e_0$. However, the sum and the product of two probabilistic observables $A,B$ are, in general, not defined. If system observables exist for arbitrary real $c_k$ this allows a definition of linear combinations of probabilistic system observables as represented by linear combinations of the associated vectors $e$. For $C=c_AA+c_BB$ one has $e^{(C)}=c_Ae^{(A)}+c_Be^{(B)}$~,~$\langle C\rangle=c_A\langle A\rangle+c_B\langle B\rangle$. 

Next we are interested in some general properties of the basis observables. For example, one typical question may ask if two of them can have simultaneously a sharp value.  A ``classical eigenstate'' of a probabilistic observable $A$ is an ensemble for which $A$ has a ``sharp value'' with vanishing dispersion, $\langle A^2\rangle -\langle A\rangle^2=0$. For example, the eigenstate of the observable $A^{(k)}$ with ''classical eigenvalue'' $\kl \A\kr=1$ is characterized by $p_\tau=0$ whenever $A^{(k)}_\tau=-1$. The maximal number of sharp ``basis observables'' $A^{(k)}$ can be characterized by the ``purity'' $P$ of the ensemble,
\be\label{3}
P=\rho_k\rho_k.
\ee
(Note that $P$ depends on the set of basis observables that characterize the subsystem.)  For $P=0$ one finds equipartition with $\langle A\rangle=0$ for all two level observables. Obviously, $\tilde M$ sharp observables require $P\geq \tilde M$, since at least for $\tilde M$ values of $k$ one needs $\rho_k=\pm 1$. For an ensemble with $P=1$ at most one observable $A^{(k)}$ can be sharp. Typical classical ensembles that describe isolated quantum systems will have a maximal purity smaller than $n$, such that not all $A^{(k)}$ can have sharp values simultaneously. We recall that the purity \eqref{3} is a statistical property involving expectation values. For a given classical state $\tau$ all observables $A^{(k)}$ have a sharp value.

\medskip
\noindent
{\bf Density matrix and wave function}

For $M$ an integer obeying $M\geq P+1$ we may represent the $\rho_k$ by an $M\times M$ hermitean ``density matrix''$\rho_{\alpha\beta}$:
\ba\label{3a}
\rho=\frac 1M(1+\rho_kL_k).
\ea
The matrices $L_k$ \cite{CW1} are $SU(M)$-generators $k=1..M^2-1$, obeying
\ba\label{3b}
\text{tr}L_k&=&0~,~L^2_k=1~,~\text{tr}(L_kL_l)=M\delta_{kl},\\
\{L_k,L_l\}&=&2\delta_{kl}+2d_{klm}L_m~,~[L_k,L_l]=2if_{klm}L_m.\nn
\ea
The matrix $\rho$ has indeed the properties of a density matrix, 
\ba
\text{tr}\rho=1~,~\rho_{\alpha\alpha}\geq 0~,~\text{tr}\rho^2=\frac 1M(1+P)\leq 1.
\ea
In analogy to quantum mechanics, a ``classical pure state'' obeys $\rho^2=\rho$ and therefore requires $P=M-1$. We can also associate to any system observable $A$ an operator $\hat A$ such that the quantum mechanical rule for the computation of expectation values holds $(e_k\equiv e^{(A)}_k=c_k)$ 
\be\label{5}
\hat A=e_kL_k~,~\langle A\rangle =\rho_ke_k=\text{tr}(\rho\hat A).
\ee
We will concentrate on the minimal $M$ needed for a given maximal purity of the ensemble. For $n=M^2-1$ the operators for the basis variables $\A$ are given by $L_k$. If $n<M^2-1$ only part of the $L_k$ are used as a basis for the observables.

Many characteristic features of the system observables $A$ can now be inferred from standard quantum mechanics, as demonstrated by a few examples. For $M=2$ at most one of the three possible two-level-observables $A^{(k)}$ can have a sharp value. This occurs for an ensemble where the density matrix describes a quantum mechanical pure state, $\rho=\frac12(1\pm \hat A)$, tr$\rho^2=\frac12+\frac14$tr$\hat A^2=1$, with $\hat A^{(k)}=L_k=\tau_k$. For such an ensemble the expectation value of the two orthogonal two-level-observables must vanish, $\langle A^{(l)}\rangle=0$ for $l\neq k$. Thus, whenever one basis observable is sharp, the two others have maximal uncertainty, as for the spin one-half system in quantum mechanics. 

Another example for $M=4$ describes two different two-level-observables (say the $z$-direction of two spins $S^1_z,S^2_z$) by $L_1=$ diag $(1,1,-1,-1)$ and $L_2=$ diag $(1,-1,1,-1)$. The product of the two spins is represented by $L_3=$ diag $(1,-1,-1,1)$. Consider an ensemble characterized by $\rho_3=-1~,~\rho_1=\rho_2=0$. For this ensemble one has $\langle A^{(1)}\rangle=\langle A^{(2)}\rangle =0$ such that for both two-level-observables the values $+1$ and $-1$ are randomly distributed in the ensemble. Nevertheless, $\langle A^{(3)}\rangle=-1$ indicates that the two ``spins'' are maximally anticorrelated. Whenever the first spin takes the value $+1$, the second one necessarily assumes $-1$ and vice versa. At the end of this note we discuss that pure states of this type show the characteristics of an entangled quantum state. Our third example considers the observable $S$ corresponding to the sum $\hat S=L_1+L_2$. For the particular pure state density matrices $(\hat\rho_m)_{\alpha\beta}=\delta_{m\alpha}\delta_{m\beta}$ one has $\langle S\rangle=2$ (for $m=1$), $\langle S\rangle=0$ (for $m=2,3)$ and $\langle S\rangle=-2$ (for $m= 4)$. Thus $S$ has the properties of a total spin, composed of two half integer spins (say $S_z=S^1_z+S^2_z)$. 

The density matrix can be diagonalized by a unitary transformation. In consequence, any pure state density matrix can be written in the form $\rho=U\hat\rho_mU^\dagger$ for a suitable $U,$ with $UU^\dagger=1$. This allows us to ``take the root'' of a pure state density matrix by introducing the quantum mechanical wave function $\psi_\alpha$ as an $M$-component complex normalized vector, $\psi^\dagger\psi=1$,
\ba\label{6}
\rho_{\alpha\beta}&=&\psi_\alpha\psi^*_\beta~,~\psi_\alpha=
U_{\alpha\beta}(\hat\psi_m)_\beta~,~\nn\\
(\hat\psi_m)_\beta&=&\delta_{m\beta}~,~\langle A\rangle=\psi^\dagger\hat A\psi.
\ea
All the usual rules for expectation values in quantum mechanical pure states apply. 

Pure states play a special role since they describe classical ensembles with minimal uncertainty for a given integer $P$.  For $M=4$ a pure state has purity $P=3$ and three different observables can have sharp values, corresponding to the maximum number of three commuting quantum mechanical operators. For $P>1,M=P+1$ and $\{L_k,L_l\}=2\delta_{kl}+2d_{klm}L_m$ the condition for a pure state, $\rho^2=\rho$ or $\rho_k[\rho_l d_{klm}-(M-2)\delta_{km}]=0$, is not automatically obeyed for all $\rho_k$ with $\rho_k\rho_k=P$. For a pure state, the ``copurity'' $C=$ tr$[(\rho^2-\rho)^2]$ must also vanish. While the purity $P$ is conserved by all orthogonal $SO(n)$ transformations of the vector $(\rho_k)$, pure states are transformed into pure states only by the subgroup of $SU(M)$ transformations. The $SU(M)$ transformations are realized as unitary transformations of the wave function $\psi$, where the overall phase of $\psi$ remains unobservable since it does not affect $\rho$ in eq. \eqref{6}. In our classical statistical description of quantum phenomena, the particular role of the classical pure states constitutes the basic origin for the unitary transformations in quantum mechanics. Just as in quantum mechanics, we can write the density matrix $\rho$ for an arbitrary ensemble as a linear combination of appropriate pure state density matrices. 

\medskip
\noindent
{\bf Algebra of probabilistic quantum observables}

The vector $e^{(A)}$ is sufficient for a determination of the expectation value of $A$ in any state of the subsystem (characterized by $\rho_k$). However, the typical observables for the subsystem are probabilistic observables and we further have to specify the probability distribution for a possible outcome of measurements for every state $\{\rho_k\}$. This is needed for a computation of expectation values $\kl A^p\kr$ of powers of $A$. We will next concentrate on ``quantum observables''. They constitute a subclass of the system observables for which the operators associated to $A^p$ are given by $\hat A^p$. For a given $M$ we consider observables with a spectrum of at most $M$ different values $\gamma_a$. We first consider a non-degenerate spectrum of $M$ different $\gamma_a$ and identify $a=\alpha$. The probabilities to find $\gamma_\alpha$ in the state $\rho_k$ of the subsystem are denoted by $w_\alpha(\rho_k)\geq 0~,~\sum_\alpha w_\alpha(\rho_k)=1~$, $\langle A\rangle=\sum_\alpha w_\alpha(\rho_k)\gamma_\alpha=\rho_ke_k$. The expectation value $\langle A\rangle=\textup{tr}(\rho\hat A)$ is invariant under a change of basis by unitary transformations, $\rho\to \rho'=U\rho U^\dagger~,~\hat A\to \hat A'=U\hat A U^\dagger$. We may choose a basis with diagonal $\hat A'=diag(\lambda_1,\dots,\lambda_M)~,~\langle A\rangle=\sum_\alpha\rho'_{\alpha\alpha}\lambda_\alpha$, suggesting that the spectrum $\gamma_\alpha$ can be identified with the eigenvalues $\lambda_\alpha$ of the operator $\hat A$, and $w_\alpha(\rho_k)=\rho'_{\alpha\alpha}$. 

Among the classical observables we therefore consider the subclass of quantum observables which generate probabilistic observables  with the property
\be\label{8A}
\kl A\kr=\sum_\alpha w_\alpha(\rho_k)\gamma_\alpha~,~w_\alpha(\rho_k)=\rho'_{\alpha\alpha}.
\ee
To each quantum observable, we can associate a hermitean quantum operator, with a spectrum of possible measurement values given by the eigenvalues $\lambda_\alpha=\gamma_\alpha$ of the operator. The classical probability for the outcome of the measurement in a given state is the corresponding diagonal element of the density matrix in a basis where $\hat A$ is diagonal. We describe the mathematical structures related to quantum observables in more detail at the end of this note, where we also deal with degenerate spectra of less than $M$ different $\gamma_a$. There we will construct an  explicit realization for the quantum observables.  

We can now consider powers of the probabilistic observable $A,~\langle A^p\rangle=\sum_\alpha w_\alpha(\rho_k)\gamma^p_\alpha$. The observable $A^p$ should belong to the observables of the subsystem, since it can be associated with $p$ measurements of $A$, multiplying the $p$ measurement results that must be identical. We can therefore associate an operator $\tilde A_p$ to the observable $A^p~,~\langle A^p\rangle=\textup{tr}(\rho\tilde A_p)$. For quantum observables this is realized by $\tilde A_p=\hat A^p$ and we conclude $\langle A^p\rangle=\textup{tr}(\rho\hat A^p)$. For $\langle A^2\rangle=$tr$(\rho\hat A^2)$ a classical eigenstate of $A$ obeys [tr$(\rho\hat A)]^2=$tr$(\rho\hat A^2)$ and the possible classical eigenvalues are the eigenvalues of the operator $\hat A$. If a pure state is an eigenstate of $A$ one has $\hat A\psi=\lambda\psi$ with $\lambda\equiv \lambda_\alpha=\gamma_\alpha$ one of the eigenvalues of $\hat A$. 

Consider now the classical ensemble of states $\tau$ which describe the subsystem together with its environment. It is straightforward to characterize the properties of the classical observables $A_\tau$ that are mapped to the probabilistic quantum observables. First, the spectrum of possible outcomes of individual measurements equals the (sharpe) values of the classical observable  $A_\tau$ in the classical states $\tau$. It consists of the eigenvalues of $\hat A$. Second, the probabilities for the states $p_\tau$ must be distributed such that $w_\alpha$ can be written as a linear combination of $\rho_k\kl A^{(k)}\kr$ for all states with $P\leq M-1$. At first sight this second requirement for a classical observable to be a quantum observable may seem rather special. However, in many circumstances the quantum observables are related to the basis observables by simple ``physical operations''. For example, the spin observable in an arbitrary direction obtains from the three basis observables $A^{(k)}(M=2)$ by rotation, such that the relation \eqref{8A} arises naturally \cite{CW2}. Other simple operations are the addition and multiplication of ``commuting observables'', as we will explain below. At this point it may be worthwhile to pause. We have selected a set of classical observables with a remarkable property: whenever the ensemble obeys a simple constraint on the purity, $P\leq M-1$, all the laws of quantum mechanics apply for these classical observables, as for example Heisenberg's uncertainty relation based on the commutator of the associated operators. 

In summary, we define the quantum observables as a subclass of the classical observables with a spectrum of at most $M$ different possible measurement values $\gamma_a$. Furthermore, the associated probabilities $w_a$ should be given by the relation \eqref{8A} which is linear in $\rho_k$, with coefficients depending only on the observable. This implies the relation \eqref{2} and defines $e^{(A)}_k$. (We may add a piece proportional to the  unit observable.) It is then also possible to compute the expectation values $\kl A^p\kr$ in terms of the information contained in the quantum state through an expression linear in the $\rho_k$. Therefore $\kl A^p\kr$ does not involve any properties of the environment - these expectation values are completely determined by the subsystem. To any quantum observable we can associate a unique vector $(e_k)$ and therefore a unique quantum operator. In turn, to each quantum operator we can also associate a unique probabilistic observable, with $w_\alpha$ given by eq. \eqref{8A}. On the level of probabilistic observables the quantum observables can therefore be fully characterized by the vector $(e_k)$. 

The correspondence between classical probabilistic observables and quantum operators allows for the introduction of a ``quantum product'' $AB$ between the probabilistic observables, which is associated to the operator product $\hat A\hat B$. Together with linear combinations, this defines an algebra for the probabilistic quantum observables. The algebra of probabilistic quantum observables is isomorphic to the algebra of quantum operators. For both, it can be expressed as operations in the space of $(e_k)$. In particular, we observe $A^2=A A$. We will see below that a particular use of the ``quantum product'' $A B$ arises from an investigation of conditional probabilities. 

In general, a probabilistic quantum observable $A$ describes an equivalence class of classical observables that all lead to the same quantum operator $\hat A$. The product $\hat A\hat B$ then induces a product structure between equivalence classes. If a representative for each class is selected one can also define $AB$ on the level of classical observables \cite{3}. We emphasize that the product $AB$ is {\em not} the classical or pointwise product $A\cdot B$ where $(A\cdot B)_\tau=A_\tau B_\tau$. 

\medskip
\noindent
{\bf Quantum correlations}

Beyond a rule for the computation of expectation values of observables, any  theory must provide a prediction for the outcome of two consecutive measurements. After a first measurement of the observable $A$ the result of a subsequent measurement of another observable $B$ is, in general, influenced by the first measurement. The measurement of $A$ has changed the ensemble and the knowledge of the observer. For simplicity we concentrate on two-level-observables, $\langle A^2\rangle=\langle B^2\rangle=1,~\hat A^2=\hat B^2=1$. The probability of finding $B=1$ after a measurement $A=1$ amounts to the {\em conditional probability} $(w^B_+)^A_+$. After the measurement $A=1$ the ensemble must be an eigenstate to the eigenvalue $\bar A=1$ - otherwise a subsequent measurement of $A$ would not necessarily yield the same value as the first one. Then $B$ is measured under this condition. 

We take here the attitude that there is only one given reality, but physicists can at best give a statistical description of it. The ``fundamental laws'' are genuinely of a statistical nature \cite{GenStat} and only establish relations within different possibilities for the history of the real world. Measuring for an observable $A$ in a given state the value $\gamma_{\bar\alpha}$ simply eliminates the other possible alternatives (which may have nonvanishing probabilities $w_{\alpha\neq \bar\alpha}$). After the measurement of $A$ it makes only sense to ask what are the outcomes of other measurements under the condition that $A$ has been measured to have the value $\gamma_{\bar\alpha}$. 

Correspondingly, the {\em measurement correlation} or {{\em conditional correlation}  $\langle BA\rangle_m$ multiplies the measured values of $A$ and $B$, weighed with the probabilities that they occur
\ba\label{7}
\langle BA\rangle_m&=&(w^B_+)^A_+w^A_{+,s}-(w^B_-)^A_+w^A_{+,s}\nn\\
&&-(w^B_+)^A_-w^A_{-,s}+(w^B_-)^A_-w^A_{-,s},
\ea
with $w^A_{\pm,s}$ the probability that $A$ is measured as $\pm 1$ in the state $s$ and $\langle A\rangle=w^A_{+,s}-w^A_{-,s}=$ tr $(\rho\hat A)$, $w^A_{+,s}+w^A_{-,s}=1$, $(w^B_+)^A_\pm +(w^B_-)^A_\pm=1$. The conditional correlation needs a specification of the conditional probabilities as $(w^B_\pm)^A_+$. For their computation we use the prescription that after the first measurement $A=1$ the density matrix $\rho_{A+}$ must describe an eigenstate of $A$, tr$(\hat A\rho_{A+})=1$. This effect of a first measurement may be called {\em state reduction}. The subsequent measurement of $B$ then involves this state,
\be\label{8}
(w^B_+)^A_+-(w^B_-)^A_+=\text{tr}(\hat B\rho_{A+}).
\ee
The relation \eqref{8} is based on the property that the quantum observable $B$ obeys eq. \eqref{2} for an arbitrary quantum state of the subsystem. Our assumption is therefore that after the measurement of $A$ the classical ensemble still describes a quantum system. This seems reasonable for appropriate measurements since otherwise the first measurement destroys the isolation of the subsystem instead of only changing its state. This assumption has far reaching consequences, however. It necessarily implies eq. \eqref{8} and excludes the option of using the classical correlation for the general description of subsequent measurements, as we will see below.

For $M=2$ the matrix $\rho_{A+}$ is unique, $\rho_{A+}=\frac12(1+\hat A)$, such that 
\be\label{9}
(w^B_\pm)^A_+=\frac12\pm \frac14\text{tr}
(\hat B\hat A)~,~(w^B_\pm)^A_-=\frac12\mp\frac14\text{tr}(\hat B\hat A).
\ee
However, for $M>2$ one has tr$(\hat A\rho_{A+})=1$ for 
\ba\label{10}
\rho_{A+}&=&\frac 1M(1+\hat A+X)~,~\text{tr}(\hat A X)=0~,~\nn\\
\text{tr}X&=&0~,~\text{tr}X^2=M(P-1),
\ea
with
\be\label{11}
\rho^2_{A+}-\rho_{A+}=\frac{1}{M^2}(X^2+\{\hat A,X\})-\left(1-\frac 2M\right)\rho_{A+}.
\ee
A necessary condition for $\rho_{A+}$ describing a pure state is tr$X^2=M(M-2)~,~P=M-1$, which implies $X=0$ only for $M=2$.  

We may distinguish between a ``maximally destructive measurement'' where all information about the original ensemble except for the value of $A$ is lost, and a ``minimally destructive measurement'' for which an original pure state remains a pure state after the measurement. A maximally destructive measurement is described by $X=0$ in eq. \eqref{10}, leading to 
\be\label{12}
\langle B\rangle_{A+}=(w^B_+)^A_+-(w^B_-)^A_+=\frac1M\text{tr}(\hat B\hat A)=\langle BA\rangle_{\max}.
\ee
Here we denote by $\langle BA\rangle_{\max}$ the conditional correlation for maximally destructive measurements and use that $\rho_{A-}$ obtains from $\rho_{A+}$ by changing the sign of $\hat A$ in eq. \eqref{10} (with $X=0$). We can use $\langle BA\rangle_{\max}$ for the definition of a scalar product between the observables $B$ and $A$, since it does not depend on the initial ensemble. The two-level observables $A^{(k)}$ form an orthogonal basis in this sense, $\langle A^{(k)}A^{(l)}\rangle_{\max}=\delta_{kl}$. 

A minimally destructive measurement of $A=1$ projects out all states with $A=-1$, without further changes of the original ensemble and associated density matrix $\rho$,
\be\label{13}
\rho_{A\pm}=\frac{1}{2(1\pm\langle A\rangle)}(1\pm\hat A)\rho(1\pm\hat A).
\ee
With
\be\label{16A}
\langle BA\rangle_m =\text{tr}(\hat B\rho_{A+})w^A_{+,s}-\text{tr}(\hat B\rho_{A-})w^A_{-,s}, 
\ee
and 
\be\label{14a}
w^A_{\pm,s}=(1\pm\langle A\rangle)/2
\ee
this yields
\be\label{14}
\kl BA\kr_m=\frac12\text{tr}(\{\hat A, \hat B\}\rho). 
\ee
The conditional correlation for minimally destructive measurements in the classical statistical ensemble corresponds precisely to the expression of this correlation in quantum mechanics. It involves the anticommutator and is therefore related to the quantum mechanical operator product. On the level of probabilistic observables we can express the conditional correlation $\kl BA\kr$ in terms of expectation values of the quantum product $AB$
\be\label{17A}
\kl BA\kr_m=\frac12(\kl BA\kr+\kl AB\kr),
\ee
demonstrating the close connection between the quantum product and the conditional correlation.

The two point correlation is commutative, $\kl BA\kr_m=\kl AB\kr_m$. We will postulate that the two point correlation \eqref{17A} describes in general the correlation between two measurements for quantum systems and call it ``{\em quantum correlation}''. We have motivated its use by two subsequent measurements, but the order of the measurements does actually not matter. Any rule for the computation of a correlation between two measurements needs some choice of an appropriate product of observables. At first sight, a possible alternative choice may be the classical correlation which is based on the classical product $A\cdot B$, as defined on the level of the classical ensemble, $(A\cdot B)_\tau=A_\tau B_\tau$. We will see below, however, that $A\cdot B$ is usually not a quantum observable and can therefore not be determined from the information characterizing a quantum state, i.e. $\{\rho_k\}$. Its use would therefore need information which relates to the environment, but not only to the subsystem. In other words, the use of the classical product corresponds to a state reduction after the first measurement where substantial information about the relation between the subsystem and the environment is retained. This is not what a usual measurement in a quantum system does and the classical correlation can therefore not serve for the description of such measurements. For any measurement where the outcome (including the state reduction) can be expressed in terms of information available for the subsystem the choice of the quantum product seems natural. It retains a maximum of the information which is available in the subsystem.

Since the correct choice of the correlation for a description of two measurements is crucial we may describe the issue in some more detail. In any statistical setting one should distinguish the probability $w_{++}$ that the two level observables $A$ and $B$ are {\em measured} with values $A=1,B=1$ from the probability $p_{++}$ that they ``have'' the values $A=B=1$ in the classical statistical ensemble. On the classical statistical level one can express
\be\label{19A}
p_{++}=\frac14(1+\kl A\kr+\kl B\kr+\kl A\cdot B\kr)
\ee
in terms of the ``classical'' correlation
\be\label{19B}
\kl A\cdot B\kr=\sum_\tau p_\tau A_\tau B_\tau.
\ee
The probability $p_{++}$ does not specify a priori the conditional information relating two subsequent measurements, which is necessary for $w_{++}$. In general, one needs a separate prescription how $w_{++}$ should be computed from the available statistical information. Only under particular circumstances, one may be able to identify $w_{++}$ with $p_{++}$. In other words, $w_{++}=p_{++}$ is an additional basic assumption which does not hold true in general. This contrasts to the case of a single measurement for $A$, where $w_+=p_+=(1+\kl A\kr)/2$ by definition.

If we try the identification $w_{++}=p_{++}$ we run into severe problems. In general, the classical correlation $\kl A\cdot B\kr$ is not computable in the subsystem. It is a property of the system and its environment and cannot be obtained from the information characterizing the quantum state, i.e. from $\rho_k$. This problem is closely linked to the fact that the mapping from the classical observables described by $A_\tau$ to the probabilistic observables described by $\gamma\Ai_a=\pm 1$ and $w\Ai_a=w\Ai_\pm$ is not invertible. The observable $A_\tau$ describes properties of the subsystem and its environment, while the characterization of the environment is only lost on the level of the probabilistic observable $A$. The use $w_{++}=p_{++}$ corresponds to an implicit definition of the conditional probability for two measurements where after the first measurement $A=1$ all classical states $\tau$ for which $A_\tau=-1$ are eliminated. This elimination process depends, however, on the particular observable $A_\tau$ and therefore also reflects properties of the environment, not only of the subsystem. Two observables $A_\tau$ and $A'_\tau$, which lead to the {\em same} probabilistic observable $A$, produce, in general, {\em different} classical products $\kl A\cdot B\kr\neq \kl A'\cdot B\kr$. (These issues are discussed in more detail at the end of this note.)

If the state of an isolated subsystem can be described by $\{\rho_k\}$, this information must also be sufficient for a prediction of the outcome of two measurements. The probability $w_{++}$ must be computable in terms of $\{\rho_k\}$. For this reason we employ the quantum correlation $\kl AB\kr$ and postulate
\be\label{19C}
w_{++}=\frac14(1+\kl A\kr+\kl B\kr+\kl AB\kr_m),
\ee
as advocated already before for subsequent measurements. (A different motivation for the use of the quantum correlation is given in \cite{3}.) Indeed, now $w_{++}$ can be expressed in terms of $\rho_k$
\be\label{19D}
w_{++}=\frac14\Big(1+e\Ai_ke\B_k+\rho_k
[e\Ai_k+e\B_k+d_{mlk}e\Ai_me\B_l]\Big)
\ee
The other probabilities $w_{+-}$ etc. obtain from eqs. \eqref{19C}, \eqref{19D} by appropriate changes of relative signs.

The prescription for the probabilities of the outcome of two measurements influences strongly the statistical properties of correlations. For example, one may ask if a ``hidden variable theory'' is possible, where there exist discrete functions $\tilde A(v)=\pm 1~,~\tilde B(v)=\pm 1$ such that $\kl AB\kr_m=\int dv\tilde p(v)\tilde A(v)\tilde B(v)$ with some probability distribution $\tilde p(v)$. Just as in the quantum mechanics, this can be excluded by the use of Bell's inequalities \cite{Bell}. In our classical statistical setting the correlation function \eqref{14} or the probabilities for the outcome of two measurements are exactly the same as in quantum mechanics. On the other hand, Bell's inequalities apply to the classical correlation $\kl A\cdot B\kr$. Besides theoretical arguments we have therefore also experimental evidence that in general the classical correlation function should not be used for the description of the outcome of two measurements.

\medskip
\noindent
{\bf Quantum time evolution}

We have seen how quantum structures can arise from the description of subsystems where the ``state of the system'' is described by $n$ expectation values of ``basis observables''. For $P<n$ the appearance of ``non-commuting structures'' is mandatory. The question remains why such quantum systems are omnipresent in nature, in contrast to ``commuting structures'' for $P=n$. The answer may be rooted in stability properties of the time evolution.

Let us consider some continuous time evolution of the classical probability distribution $\{p_\tau\}$. It relates the ensemble at time $t_2$ to the ensemble at some earlier time $t_1$, and induces a transition from $\rho_k(t_1)$ to $\rho_k(t_2)$, 
\be\label{15}
p_\tau(t_2)=\tilde S_{\tau\rho}(t_2,t_1)p_\rho(t_1)~,~\rho_k(t_2)=S_{kl}(t_2,t_1)\rho_l(t_1).
\ee
We may decompose the transition matrix $S_{kl}$ into the product of an orthogonal matrix $\hat S_{kl}$, which preserves the length of the vector $(\rho_1\dots,\rho_n)$ and therefore the purity, and a scaling $d,~S_{kl}=\hat S_{kl}d$. For an infinitesimal evolution step this implies
\ba\label{16}
\partial_t\rho_k(t)&=&T_{kl}\rho_l(t)+D\rho_k(t)~,~D=\partial_t\ln d(t,t_1),\nn\\
T_{kl}&=&-T_{lk}=\partial_t\hat S_{km}(t,t_1)\hat S_{lm}(t,t_1).
\ea
For a given maximal purity during the evolution, eq. \eqref{16} can be rewritten as an equation for the density matrix $\rho$,
\ba\label{17}
\partial_t\rho_{\alpha\beta}=&-&i[H,\rho]_{\alpha\beta}+R_{\alpha\beta\gamma\delta}
\left(\rho_{\gamma\delta}-\frac 1M\delta_{\gamma\delta}\right)\nn\\
&+&D(\rho_{\alpha\beta}
-\frac1M\delta_{\alpha\beta}).
\ea
This corresponds to a split of the infinitesimal $SO(n)$ transformation $\delta\rho_k=T_{kl}\rho_l$ into a unitary part corresponding to the subgroup $SU(M)$ and represented by the hermitean Hamiltonian $H=H_kL_k+H_0$, and remaining rotations of $SO(n)/SU(M)$ represented by $R$ or $\tilde T_{kl}$, 
\ba\label{18}
T_{kl}&=&-2f_{klm}H_m+\tilde T_{kl}~,~[L_k,L_l]=2if_{klm}L_m,\nn\\
\tilde T_{kl}&=&\frac1M
R_{\alpha\beta\gamma\delta}(L_k)_{\beta\alpha}(L_l)_{\gamma\delta}. 
\ea
In general, $H,R$ and $D$ may depend on $\rho_k$. 

We are interested in possible partial fixed points of the evolution for which $R=0$ and $D=0$, while $H$ is independent of $\rho_k$. (Partial fixed points of this type have been found explicitly in the classical time evolution of non-relativistic boson fields \cite{4}.) Then eq. \eqref{17} reduces to the linear von-Neumann equation for the density matrix. In case of a pure state density matrix this implies the Schr\"odinger equation $i\partial_t\psi=H\psi$. One recovers the unitary time evolution of quantum mechanics. The more general evolution equation away from the ``unitary partial fixed point'' can describe ``decoherence'' \cite{DC} as a decrease of purity for $D<0$, or ``syncoherence'' as the approach to the pure state partial fixed point with increasing purity for $D>0$. The latter typically accounts for a situation where the subsystem described by the observables $A^{(k)}$ can exchange energy with the environment. An example is the evolution from a mixed state of atoms in different energy states to a pure state of atoms in the ground state by virtue of radiative decay of the excited states. A static pure state density matrix obtains as usual as a solution of the quantum mechanical eigenvalue problem $H\psi=E_j\psi$. We suggest that the omnipresence of quantum systems in nature is due to the existence of such partial fixed points which reflect the isolation of the subsystem. 

If $H$ is independent of $\rho_k$ it can be considered as an observable of the subsystem, $H=H_kL_k+H_0$, with fixed coefficients $H_k,H_0$. By Noether's theorem it is associated with the energy of the subsystem, where $E_j$ denotes the possible energy eigenvalues. (If one wants to use standard energy units one replaces $H\to H/\hbar$.) On the other hand, $R$ and $D$ account for the interactions of the subsystem with its environment. They vanish in the limit of ``perfect isolation'' of the subsystem. If the interactions with the environment are strong enough, the subsystem is typically not evolving towards the equipartition fixed point,  $\rho_{\alpha\beta}=\frac1M\delta_{\alpha\beta}$, but rather towards a Boltzmann type density matrix $\rho\sim \exp[-\beta(H+\mu_i N_i)]$ (for conserved quantities $N_i$ and chemical potentials $\mu_i$), which may be close to a pure state density matrix if the temperature $T=\beta^{-1}$ is small as compared to the typical separation of the two lowest energy eigenvalues $E_j$. In contrast, if the isolation from the environment becomes efficient fast enough, the subsystem follows subsequently its own unitary time evolution, as well known from quantum mechanics. 

At this point we may recapitulate what we have achieved. Starting form a classical statistical ensemble we have identified a class of classical observables which have the same expectation values and admit the same algebra as the operators in a corresponding $M$-state quantum system. This holds provided the ``purity condition''  $P\leq M-1$ is obeyed. Also the conditional correlations which describe measurements of two such observables are the same as in the quantum system. Furthermore, we have found the criteria for the time evolution of the classical probability distribution that ensure a unitary evolution of the density matrix for the corresponding quantum system. Obviously, these properties are related to the specific subset of classical observables that can describe the subsystem, in the sense that no further information about the environment is needed for a prediction of the outcome of measurements in the subsystem. In the remaining part of this paper we will discuss the properties of these specific quantum observables in some detail.

\medskip\noindent
{\bf Quantum observables}

The quantum structures for probabilistic observables discussed so far do not need any specification of the representation as classical observables. In eq. \eqref{1} many classical systems with different states $\tau$, classical probabilities $p_\tau$ and classical values of the observable in a given state, $A^{(C)}_\tau$, may describe the same quantum properties of the subsystem. Nevertheless, the implementation of quantum observables implies certain restrictions on the possible classical realizations. We will discuss those in the following, mainly for the purpose of conceptual foundations.

Classical observables are maps from the set of probability distributions $\Omega$, with elements $\{p_\tau\}=(p_1\dots,p_z)$, to real numbers, $\Omega{\stackrel {A^{(C)}}{\to}}{\mathbbm R}$, with $\{p_\tau\}\to\kl A^{(C)}\kr=\sum_\tau p_\tau A^{(C)}_\tau$. (For simplicity we employ a language with a finite number $z$ of classical states, which can be extended to an infinite set at the end in some specified limiting procedure.) We will restrict the discussion to those elements of $\Omega$ which correspond to ``quantum states'', i.e. which obey the bound for the purity of the ensemble. In general, the classical observables $A\C$  describe the system and its environment. (We use ``system'' for the (isolated) subsystem or quantum system from now on.) 

We are interested in the subclass of quantum observables $A^{(Q)}$ whose expectation values and quantum correlations can be computed in the system. Quantum observables are classical observables with special properties. First, for a $M$-state quantum system, a quantum observable has at most $M$ different classical values $A^{(Q)}_\tau=\gamma_a$. (For a non-degenerate spectrum we can identify $a=\alpha=1\dots M.)$ For any given quantum observable we can classify the classical states $\tau$ according to the value of $A^{(Q)}_\tau~,~\tau=(\gamma_a,\sigma_{\gamma_a})$, where for each given $\gamma_a$ one typically has a large degeneracy of classical states, labeled by $\sigma_{\gamma_a}$. We can define the probability $w_a$ for the occurrence of a possible measurement value $\gamma_a$ as 
\be\label{W1}
w_a=\sum_{\sigma_{\gamma_a}} p_{(\gamma_a,\sigma_{\gamma_a})}~,~\kl A^{(Q)}\kr=\sum_a\gamma_a w_a.
\ee

As a crucial ingredient, $w_a$ must be computable from the quantities which specify the quantum state, i.e. from the expectation values of the ``basis observables'' 
$\kl A^{(k)}\kr=\rho_k$. We assume a linear relation 
\be\label{W2}
w_a{(\rho_k)}=\sum_{k} c_{ak}\rho_k+c_{a0}.
\ee
The probabilities must be normalized, $\sum_aw_a(\rho_k)=1$, for arbitrary $\rho_k$, which implies the conditions
\be\label{30X}
\sum_ac_{ak}=0~,~\sum_ac_{a0}=1.
\ee
For quantum observables the coefficients $c_{ak},c_{a0}$ are restricted further since $w_a(\rho_k)$ must obey eq. \eqref{8A}. For non-degenerate eigenvalues one needs
\be\label{30A}
c_{\alpha k}=\frac1M (UL_kU^\dagger)_{\alpha\alpha}~,~c_{\alpha 0}=\frac1M,
\ee
for some suitable unitary matrix $U$. 

Thus quantum observables are defined by two properties:\\
(i) the restriction of the spectrum to at most $M$ different values, (ii) the  ``quantum determination'' of probabilities $w_a(\rho_k)$. While (i) only involves a property of the classical observable, the second restriction (ii) depends on the selection of possible quantum states out of the most general probability distributions $\{p_\tau\}$ and on relations to the basis observables. For the specification of a quantum observable we need at least $\gamma_a$ and $c_{ak},c_{a0}$. A quantum observable $A^{(Q)}$ has the important property that its classical product $A^{(Q)}\cdot A^{(Q)}$ (defined by $(A^{(Q)}\cdot A^{(Q)})_\tau=(A^{(Q)}_\tau)^2)$ is again a quantum observable, with spectrum $(\gamma^2_a)$ and the same $w_a$ as for $A^{(Q)}$. This extends to higher polynomials and arbitrary functions $f(A^{(Q)})$.

We can now associate to any classical quantum observable $A^{(Q)}$ a probabilistic quantum observable, $A^{(Q)}\to A$, which is characterized by the spectrum of possible measurement values $(\gamma_a)$ and the associated probabilities $w_a$. Only this information will be needed for a computation of expectation values $\kl(A^{(Q)})^p\kr$ in a ``quantum state'' of the system, while the detailed form of $\{p_\tau\}$ is not relevant. On the level of classical observables the quantum observables are characterized by a distribution of values $A\Q_\tau=\gamma_a$ for the classical states $\tau$. This distribution still contains much more information than the spectrum $\gamma_a$ and the associated probabilities $w_a$. Therefore $A\Q_\tau$ still describes the system and partly the environment. Only on the level of probabilistic observables $A$ the parts of $A\Q_\tau$ relevant for the environment are projected out, such that $A$ only ``measures'' properties of the system.

We also can associate to every $A^{(Q)}$ a quantum operator $\hat A$ by a map $A^{(Q)}\to \hat A$. It is constructed from eq. \eqref{5} by observing
\ba\label{W3}
\kl A^{(Q)}\kr&=&\sum_a\gamma_a(c_{a,0}+\sum_k c_{a,k}\rho_k)
=e^{(A)}_0+\sum_k e^{(A)}_k\rho_k,\nn\\
e^{(A)}_0&=&\sum_a\gamma_ac_{a,0}~,~e^{(A)}_k=\sum_a\gamma_a c_{a,k},\nn\\
\hat A&=&e^{(A)}_0+\sum_k e^{(A)}_kL_k~,~\kl A\Q\kr=\tr(\hat A\rho).
\ea
We note that the map \eqref{W3} is possible for arbitrary probabilistic system observables obeying eq. \eqref{W2}. Without the restriction of the type \eqref{30A} for quantum observables, however, $(A\Q)^2$ will, in general, not be mapped to $\hat A^2$. A simple example is the random two-level observable $R$, with $\gamma_1=1~,~\gamma_2=-1~,~c_{ak}=0~,~c_{10}=c_{20}=1/2$. It is mapped to $\hat A=0$, while $R^2=1$. 

In turn, we have a map from the space of quantum operators ${\cal O}$ to the space of probabilistic observables ${\cal P}$, since for every operator $\hat A$ the spectrum $\{\gamma_a\}$ is defined, and the probabilities $w_a$ can be computed for all quantum states $\{\rho_k\}$ or density matrices $\rho$. The latter obtain from the diagonal elements $(U\rho U^\dagger)_{\alpha\alpha}$, with $U$ the unitary matrix used for the diagonalization of $\hat A$. 

However, the map from the ensemble of classical quantum observables ${\cal Q}$ to the space of quantum operators ${\cal O}$ is not invertible. The classical observables $A^{(Q)}$ involve a specification of $A^{(Q)}_\tau$ for every classical state $\tau$, which is much more information than contained in the coefficients $c_{a,k}~,~c_{a,0}$. We may encounter situations where a quantum observable $B$ is mapped to an operator $\hat B$, while also a function $f(A)$ of a different quantum observable $A$ is mapped to the same operator, $f(\hat A)=\hat B$. (Here $f(\hat A)$ is an operator valued function, while $f(A)$ is based on the classical product $A\cdot A$.) Such a situation does not imply an identification of the quantum observables, i.e. in general one has $B\neq f(A)$. This lack of invertibility of the map ${\cal Q}\to {\cal O}$ constitutes an important difference between our approach and many alternative attempts of a ``classical formulation of quantum mechanics'', which associate to each $\hat A$ a unique classical observable. For example, this is typically assumed for ``hidden variable theories''. Also for the Kochen-Specker theorem \cite{KS} the existence of a map $\hat A\to A^{(Q)}$ is a crucial hypothesis, which is not obeyed in our setting.

On the level of classical observables we always can define a linear combination, $C=\lambda_AA^{(Q)}+\lambda_BB^{(Q)}$, and the pointwise product, $D=A^{(Q)}\cdot B^{(Q)}$, of two quantum observables $A^{(Q)},B^{(Q)}$, where $C_\tau=\lambda_AA^{(Q)}_\tau+\lambda_BB^{(Q)}_\tau,D_\tau=A^{(Q)}_\tau B^{(Q)}_\tau$. However, in general neither $C$ nor $D$ are quantum observables. Consider the simplest case, $M=2$, and the two basis observables $A^{(1)}$ and $A^{(2)}$ with spectrum $\gamma^{(1)}_\alpha=\gamma^{(2)}_\alpha=\pm 1$. A linear combination $C=\cos \vartheta A^{(1)}+\sin\vartheta A^{(2)}$ has a spectrum $\gamma\C_\alpha=\pm \cos \vartheta\pm \sin\vartheta$. This observable has four different possible measurement values. It can therefore not be a quantum observable of the system with $M=2$, even though $\kl C\kr$ can be computed in terms of $\rho_{1,2}=\kl A^{(1),(2)}\kr$. We conclude that the ``rotated spin'', which corresponds to the operator $\hat A(\vartheta)=\cos \vartheta\hat A^{(1)}+\sin\vartheta\hat A^{(2)}$, has to be described by a quantum observable $A^{(Q)}_{(\vartheta)}$ that is again a two level observable with spectrum $\gamma_\alpha=\pm 1$, rather than by a linear combination of $A^{(1)}$ and $A^{(2)}$ of the type $C$. This necessity arises for each  value of the angle $\vartheta$ and we have discussed in detail in \cite{CW2} that this needs a classical ensemble with infinitely many classical states $\tau$. The reader should note that one can define two types of linear combinations. On the level of classical observables one can define combinations of the type $C$, while on the level of operators or the associated quantum observables a natural definition is $\hat A(\vartheta)$ or $A^{(Q)}(\vartheta)$. In general, a projection on the subsystem does not map $C$ to $\hat A(\vartheta)$ or the probabilistic observable $A^{(Q)}(\vartheta)$.

The classical product $D=A^{(1)}\cdot A^{(2)}$ has a spectrum $\gamma^{(D)}_\alpha=\pm 1$. The condition (i) for a quantum observable is obeyed by $D$. However, the probability $w^{(D)}_\alpha$ for finding $\gamma^{(D)}_\alpha=1$ needs knowledge of the joint probability to find $A^{(1)}=1,A^{(2)}=1$ or $A^{(1)}=-1,~A^{(2)}=-1$. This information cannot be extracted from $\rho_1$ and $\rho_2$, which only yield the probabilities for finding $A^{(1)}=\pm 1$ (namely $w^{(1)}_\pm =(1\pm \rho_1)/2)$ or for finding $A^{(2)}=\pm 1$ (namely $w^{(2)}_\pm=(1\pm\rho_2)/2)$. Nor is it contained in the expectation value $\rho_3$ of the third basis variable for $M=2$. We conclude that the classical observable $D$ does not obey the condition (ii) for a quantum observable. 

\bigskip
\noindent
{\bf Representation as classical observables}

After these general remarks we now present an explicit classical ensemble and quantum observables for a system with given $M$. We recall that the classical quantum observables $A^{(Q)}$ which are mapped to a given $\hat A$ are not unique. Also the specification of the classical states $\tau$ and the corresponding construction of classical observables is not supposed to be unique. At the end all measurable information of the system can be expressed in terms of the expectation values of quantum operators, such that the details of the classical observables do not matter. Only the existence of the classical observables in a setting free of contradictions is therefore needed in order to demonstrate a realization of quantum mechanics as a classical statistical ensemble. 

It is sufficient to determine at least one classical quantum observable for every operator $\hat A$, i.e. for every hermitean $M\times M$ matrix. The quantum observable $\lambda A^{(Q)}$, with $\lambda\in {\mathbbm R}$, is mapped to the operator $\lambda\hat A$. We will therefore restrict our discussion to operators with unit norm, say tr $\hat A^2=M$. Also the addition of a part proportional to the unit observable translates for operators to the addition of a corresponding piece proportional to the unit operator, $A+c\to \hat A+c$. We can therefore restrict the discussion to traceless operators, tr$\hat A=0$. We follow a simple construction principle. Consider first a single operator $\hat A$ with a spectrum of $m(\hat A)\leq M$ distinct eigenvalues $\lambda_{a(\hat A)}(\hat A)$. We associate to it $m(\hat A)$ discrete classical states, labeled by $a(\hat A)=1\dots m(\hat A)$. In these states the classical observable $A^{(Q)}$, which is mapped to $\hat A$, takes the values $A^{(Q)}_{a(\hat A)}=\lambda_{a(\hat A)}(\hat A)$. Add now a second operator $\hat B$ with $m(\hat B)$ distinct eigenvalues $\lambda_{a(\hat B)}(\hat B)$. If this operator is ``independent'' we construct the direct product space with states $\tau$ labeled by the double index $\tau=\big(a(\hat A),a(\hat B)\big)$, and $A^{(Q)}_\tau=\lambda_{a(\hat A)}(\hat A),B^{(Q)}_\tau=\lambda_{a(\hat B)}(\hat B)$. This is continued until all independent operators are included. As stated above, the resulting ensemble has infinitely many classical states $\tau$, since the number of independent operators is infinite. (A well defined sequence of subsequently included operators induces a well defined limit process for the construction of the ensemble \cite{CW2}.) Our construction yields explicitly a classical quantum observable for every independent operator. We recall that many further classical observables that do not obey the restrictions for quantum observables can be defined in the ensemble.

Details of the construction will depend on the notion of independent operators. As a simple criterion we may call two operators independent if tr$(\hat A-\hat B)^2\geq\epsilon$, and take the limiting process $\epsilon\to 0$. Other more restrictive definitions of ``independent'' may be possible. Different contradiction-free definitions of ``independent'' lead to different classical realizations, which all result in the same quantum properties of the system. The explicit construction above has demonstrated that such classical realizations exist. Of course, there are also classical ensembles with ``many more'' states than those used in our explicit construction. If one is only interested in the quantum observables the states $\tilde \tau$ of such a larger ensemble can be mapped to the states $\tau$ of the ensemble used in the construction by summing the probabilities of all states which have the same $A_\tau$ for all quantum observables. 

\medskip
\noindent
{\bf Classical product for observables}

We next turn to the conditions under which linear combinations or classical products of two quantum observables are again quantum observables. For $C^{(Q)}=\lambda_AA^{(Q)}+\lambda_BB^{(Q)}$ the expectation value can always be expressed in terms of $\rho_k$. We can therefore compute $e^{(C)}_k=\lambda_Ae^{(A)}_k+\lambda_Be^{(B)}_k$ such that $\kl C\kr$ obeys eq.\eqref{2}. This allows for the construction of an operator $\hat C$ \eqref{5} which obeys for all $\rho$
\be\label{W4}
\kl C\kr=\textup{tr}(\hat C\rho)=\lambda_A\kl A\kr+\lambda_B\kl B\kr=\textup{tr}
\Big[(\lambda_A\hat A+\lambda_B\hat B)\rho\Big].
\ee
We therefore can identify $\hat C=\lambda_A\hat A+\lambda_B\hat B$. At this step, however, the spectrum of possible measurement values for $C$ does not necessarily coincide with the spectrum of eigenvalues of $\hat C$, which is a necessary condition for a quantum observable. On the classical level the spectrum of $C$ consists of all linear combinations
\be\label{Z1}
\gamma\C_c=\gamma\C_{(a,b)}=\lambda_A\gamma^{(A)}_a+\lambda_B\gamma^{(B)}_b
\ee
for all possible pairs $(a,b)=c$. It may be reduced by those $\gamma\C_c$ for which the probability $w_c$ vanishes for all quantum states. Even if the number $\tilde M\C$ of different values $\gamma\C_c$ obeys $\tilde M\C\leq M$ as realized, for example, if $\tilde M\Ai\tilde M\B\leq M$, there is no guarantee that all $\gamma\C_c$ coincide with the eigenvalues of $\hat C$.

If $C\Q$ is a quantum observable its spectrum must coincide with the spectrum of $\hat C$. Furthermore, the classical product $C\Q\cdot C\Q$ must also be a quantum observable and obey
\ba\label{Z2}
\kl C\cdot C\kr&=&\lambda^2_A\kl A\cdot A\kr+\lambda^2_B\kl B\cdot B\kr +
2\lambda_A\lambda_B\kl A\cdot B\kr\nn\\
&=&\textup{tr}(\hat C^2\rho)\\
&=&\lambda^2_A\textup{tr}(\hat A^2\rho)+\lambda^2_B\textup{tr}(\hat B^2\rho)+
\lambda_A\lambda_B\textup{tr}\Big(\{\hat A,\hat B\}\rho\Big).\nn
\ea
We find as a necessary condition that the classical product $A\cdot B$ must be computable in terms of $\rho_k$
\be\label{Z3}
\kl A\cdot B\kr=\frac12\tr \Big(\{\hat A,\hat B\}\rho\Big).
\ee
This has to hold for arbitrary $\rho$. If any (nontrivial) linear combination of $A\Q$ and $B\Q$ is a quantum observable one concludes that $A\Q\cdot B\Q$ must be a quantum observable with associated operator $A\cdot B\to \frac12\{\hat A,\hat B\}$. Similar restrictions arise for higher powers of $C$. 

The condition \eqref{Z3} is nontrivial. On the classical level we can derive from $\{p_\tau\}$ the probabilities $w_{(a,b)}$ that $A$ has the value $\gamma\Ai_a$ and $B$ takes the value $\gamma\B_b$. For quantum observables $A\Q,B\Q$ the probabilities $w\Ai_a=\sum_b w_{(a,b)},w\B_b=\sum_a w_{(a,b)}$ can be computed from $\{\rho_k\}$. In general, the information contained in $\{\rho_k\}$ will not be sufficient to determine $w_{(a,b)}$, however. It will therefore often not be possible to express $\kl A\cdot B\kr=\sum_{a,b}\gamma\Ai_a\gamma\B_b w_{(a,b)}$ in terms of $\{\rho_k\}$. Then $D=A\cdot B$ cannot be a quantum observable. On the other hand, if a linear relation between $\kl D\kr$ and $\rho_k$ exists, $\kl D\kr=e\iD_0+e\iD_k\rho_k$, we can write $\kl D\kr=\tr(\hat D\rho)$ and eq. \eqref{Z3} implies $\hat D=\frac12\{\hat A,\hat B\}$. 

If $\lambda_AA\Q+\lambda_BB\Q$ is a quantum observable for arbitrary $\lambda_A,\lambda_B$ the associated operators $\hat A$ and $\hat B$ must commute, $[\hat A,\hat B]=0$. In order to show this, we first consider the case where a given 
$\gamma\C_{\bar c}=\lambda_A\gamma\Ai_{\bar a}+\lambda_B\gamma\B_{\bar b}$ corresponds to a unique combination $(\bar a,\bar b)$. If $C\Q$ is a quantum observable, there must exist probability distributions $\{p_\tau\}$ which are an ``eigenstate'' for the ``eigenvalue'' $\gamma\C_{\bar c}$. This implies $w_{\bar c}=1~,~w_{c\neq \bar c}=0$ or $w_{(\bar a,\bar b)}=1~,~w_{(a,b)}=0$ if $a\neq \bar a$ or $b\neq \bar b$ and therefore $w_{\bar a}=1~,~w_{a\neq\bar a}=0~,~w_{\bar b}=1~,~w_{b\neq\bar b}=0$. We conclude that this state is also a simultaneous eigenstate of the observables $A\Q$ and $B\Q$, with respective eigenvalues $\gamma\Ai_{\bar a}$ and $\gamma\B_{\bar b}$. In particular, we may consider a pure state $\psi_{\bar c}$ which is an eigenstate of $\hat C$ with eigenvalue $\gamma\C_{\bar c}$. It must obey $\hat A\psi_{\bar c}=\gamma\Ai_{\bar a}\psi_{\bar c}~,~\psi^T_{\bar c}\hat A=\gamma\Ai_{\bar a}\psi^T_{\bar c}$~,~$\hat B\psi_{\bar c}=\gamma\B_{\bar b}\psi_{\bar c}$~,~$\bar\psi^T_c\hat B=\gamma\B_{\bar b}\bar\psi^T_c$. We choose a basis where $\hat C$ is diagonal and $\psi^T_{\bar c}=\hat\psi^T_1=(1,0,\dots 0)$. In this basis $\hat A$ and $\hat B$ must be block-diagonal, $\hat A_{\alpha 1}=\gamma\Ai_{\bar a}\delta_{\alpha 1}$, $\hat A_{1\beta}=\gamma\Ai_{\bar a}\delta_{1\beta}$ and similar for $\hat B$. We can repeat this for other eigenvalues of $\hat C$. If for every eigenvalue $\gamma\C_c$ of $\hat C$ the composition out of eigenvalues of $A\Q$ and $B\Q$ is unique, we can infer that $\hat A$ and $\hat B$ must commute. Indeed, in the basis where $\hat C$ is diagonal both $\hat A$ and $\hat B$ must be simultaneously diagonal, and therefore $[\hat A,\hat B]=0$. 

In presence of multiple possibilities of composing $\gamma\C_{\bar c}$ from linear combinations of $\gamma\Ai_a$ and $\gamma\B_b$ the discussion is more involved. This case appears, however, only for particular coefficients $\lambda_A,\lambda_B$. If $C$ is a quantum observable for arbitrary $\lambda_A$ and $\lambda_B$ such degenerate cases can be avoided such that $\hat A$ and $\hat B$ must commute. Indeed, consider the case of a ``degenerate decomposition'' for a particular pair $(\lambda_A,\lambda_B)$. This occurs if there are two solutions $\lambda_A\gamma\Ai_{a_1}+\lambda_B\gamma\B_{b_1}=\gamma\C_{c_1}~,~\lambda_A\gamma\Ai_{a_2}+\lambda_B\gamma\B_{b_2}=\gamma\C_{c_2}$, with $\gamma\C_{c_1}=\gamma\C_{c_2}$~,~$\gamma\Ai_{a_1}\neq\gamma\Ai_{a_2}$. Performing an infinitesimal shift $\lambda_A\to\lambda_A+\delta_A$, while keeping $\lambda_B$ fixed, results in a separation of $\gamma\C_{c_1}$ and $\gamma\C_{c_2}$ , $\gamma\C_{c_2}-\gamma\C_{c_1}=\delta_A(\gamma\Ai_{a_2}-\gamma\Ai_{a_1})\neq 0$. Then $\gamma\C_{c_1}$ has a unique composition from $\gamma\Ai_{a_1}$ and $\gamma\B_{b_1}$. (We have discussed here the case of two-fold degeneracy where other eigenvalues of $C$ are separated from $\gamma\C_{c_1}~,~\gamma\C_{c_2}$ by a finite distance. Higher degeneracies can be treated similarly.) 

Two classical quantum observables $A\Q$ and $B\Q$ are called ``{\em commuting observables}'' if arbitrary linear combinations $\lambda_AA\Q+\lambda_BB\Q$ are also quantum observables. The operators $\hat A,\hat B$ associated to a pair of commuting observables must commute, $[\hat A,\hat B]=0$. Furthermore, the classical product of two commuting observables is a quantum observable. The associated operators and probabilistic observables are given by the chain of maps
\be\label{Z4}
A\Q\cdot B\Q\to \frac12\{\hat A,\hat B\}\to A B. 
\ee
For commuting quantum observables the classical correlation is computable from the quantum system and equals the quantum correlation. 

We next consider general conditions for the classical product being a quantum observable, $D\Q=A\Q\cdot B\Q$. If for all eigenvalues of $D\Q$ the decompositions $\gamma\iD_{\bar c}=\gamma\Ai_{\bar a}\gamma\B_{\bar b}$ are unique, the operators $\hat A$ and $\hat B$ must again commute. In this situation the quantum state specifies the probabilities $w_a$ for the observable $A$ to have the value $\gamma\Ai_a$, the analogue for $w_b$, and in addition the joint probability $w_c=w_{(a,b)}$ that a measurement of $A$ yields $\gamma\Ai_a$ and a measurement $B$ yields $\gamma\B_b$. Since the associated operators $\hat A$ and $\hat B$ commute we may choose a basis where both are diagonal. The probability for the operator $\hat A\hat B$ to take the value $\gamma\iD_c=\gamma\Ai_a\gamma\B_b$, with $(a,b)=c$, is given by the corresponding diagonal element of the density matrix in this basis, $(\rho')_{\alpha\alpha}$. This must equal $w_c$, and we conclude that for all quantum states Tr$(\hat A\hat B\rho)=\sum_cw_c\gamma\iD_c=\kl D\kr=$Tr$(\hat D\rho)$ and therefore  eq. \eqref{Z4} applies. 

\medskip
Inversely, we cannot infer that for every pair of quantum observables $A\Q,B\Q$, for which the associated operators commute, $[\hat A\Q,\hat B\Q]=0$, the classical product $A\cdot B$ must be a quantum observable. There is simply no guarantee that the joint probabilities $w_{(a,b)}$ find an expression in terms  of $\{\rho_k\}$. As an upshot of this discussion we conclude  that the lack of a map ${\cal O}\to {\cal Q}$ leaves a lot of freedom in the choice and properties of the quantum observables. Generically, linear combinations and classical products of quantum observables are not quantum observables themselves. 

\medskip
\noindent
{\bf Bit chains}

Finally, we discuss the special setting of ``bit chains''. The simplest bit chain is a set of three commuting two level observables which we take among the set of basis observables $\A$. For the example $M=4$ we may consider the observables $T_1,T_2,T_3$ associated to the three commuting diagonal operators $L_1,L_2,L_3$. A bit chain arises if the expectation values of two  (or several) one-bit-observables as well as their products can be determined simultaneously in a quantum system. Suppose that the first bit corresponds to $T_1$, the second to $T_2$. Each bit can take the two values $+1$ or $-1$. For a bit chain the classical product $T_1\cdot T_2$ is also a quantum observable, associated with $T_3$. From the expectation values $\kl T_1\kr~,~\kl T_2\kr~,~\kl T_3\kr$ we can determine all probabilities for the four possibilities of values $(+,+)~,~(+,-)~,~(-,-)$ and $(-,+)$ for bits one and two \cite{CW1}. This can be easily generalized: for a bit chain of quantum observables with $\tilde P$ members $T_j~,~j=1\dots \tilde P$, all mutual classical products are members of the bit chain, $T_i\cdot T_j=c_{ijk}T_k(i\neq j)$, with $c_{ijk}=c_{jik}=1$ for one particular combination $(i,j,k)$ and zero otherwise. All associated operators $\hat T_j$ mutually commute, and $\hat T_i\hat T_j=c_{ijk}\hat T_k~,~\hat T^2_j=1$. Also all linear combinations $\lambda_iT_i+\lambda_j T_j$ are quantum observables, represented by the operators $\lambda_i\hat T_i+\lambda_j\hat T_j$.

For given $M$ the maximal number of members of a bit chain is $\tilde P=M-1$ and we call such chains ``complete bit chains''. This restriction follows simply from the maximal number of mutually commuting operators. The presence of a bound for $\tilde P$ poses certain restrictions on the classical realizations of bit chains associated to different sets of mutually commuting operators. 

As an example, consider the case $M=8$. A possible complete three bit chain with seven members can be associated to the operators $C_1\to (\tau_3\otimes 1\otimes 1)~,~C_2\to (1\otimes\tau_3\otimes 1)~, ~C_3\to (1\otimes 1\otimes\tau_3)~,~\tilde C_1\to (1\otimes\tau_3\otimes\tau_3)~,~\tilde C_2\to(\tau_3\otimes 1\otimes\tau_3)~,~\tilde C_3\to(\tau_3\otimes\tau_3\otimes 1)~,~{\stackrel \approx C}\to (\tau_3\otimes\tau_3\otimes\tau_3)$. For this ``$C$-chain'' one has $C_j\tilde C_j={\stackrel \approx C}=C_1C_2C_3$ for all $j=1,2,3~,~\tilde C_1\cdot \tilde C_2=\tilde C_3$. Alternative candidates for complete three bit chains are the ``$A$-chain'' where $\tau_3$ is replaced by $\tau_1$, or the ``$B$-chain'' which obtains from the $C$-chain by the replacement $\tau_3\to \tau_2$, Further candidates are the ``$F$-chain'' $(C_1,A_2,A_3,\tilde F_1,\tilde F_2,\tilde F_3,{\stackrel \approx F})$, ``$G$-chain'' $(A_1,C_2,A_3,\tilde G_1,\tilde G_2,\tilde G_3,{\stackrel \approx G})$ or ``$H$-chain'' $(A_1,A_2,C_3,\tilde H_1,\tilde H_2,\tilde H_3,{\stackrel \approx H})$, with analogous multiplication structures given by the order of the elements in the list, i.e. ${\stackrel \approx F}\to (\tau_3\otimes\tau_1\otimes\tau_1)$ , ${\stackrel\approx G}=(\tau_1\otimes\tau_3\otimes\tau_1)$, ${\stackrel \approx H}\to(\tau_1\otimes\tau_1\otimes\tau_3)$. Finally, we may consider a possible candidate ``$Q$-chain'' $({\stackrel \approx F},{\stackrel\approx G},{\stackrel\approx H},\tilde Q_1,\tilde Q_2.\tilde Q_3,{\stackrel\approx Q})$, with ${\stackrel\approx Q}\to-(\tau_3\otimes\tau_3\otimes\tau_3)$. If all these sets of observables are simultaneously realized as bit chains, we run into contradiction. From the $Q$-chain we conclude ${\stackrel\approx F}\cdot{\stackrel\approx G}\cdot{\stackrel\approx H}={\stackrel\approx Q}$. In turn, from the $F,G,H$ chains we infer $C_1\cdot A_2\cdot A_3={\stackrel\approx F}$, $A_1\cdot C_2\cdot A_3={\stackrel\approx G}$ , $A_1\cdot A_2\cdot C_3={\stackrel\approx H}$ and therefore ${\stackrel\approx Q}=C_1\cdot A_2\cdot A_3\cdot A_1
\cdot C_2\cdot A_3\cdot A_1\cdot A_2\cdot C_3=C_1\cdot C_2\cdot C_3={\stackrel \approx C}$. The operators associated to ${\stackrel\approx C}$ and ${\stackrel\approx Q}$ have opposite sign, however, showing the contradiction. 

This clearly demonstrates that not every set of two level observables for which the associated operators mutually commute can be a bit chain simultaneously. In our case this only poses a consistency condition for the possibilities of classical products of quantum observables being quantum observables themselves. If we had a map $\hat A\to A\Q$, with $f(\hat A)\to f(A\Q)$ and $f(A\Q)$ based on the classical product, one could show that for every pair of commuting operators $\hat A,\hat B$ the map implies $\hat A\hat B\to A\cdot B$. The resulting contradiction is a proof of the Kochen-Specker-theorem \cite{KS} - actually the above chains of observables correspond precisely to the elegant proof of this theorem by N. Straumann \cite{Str}.

The observation that not all ``candidate chains'' $C,F,G,H,Q$ can be simultaneously bit chains does not mean that the associated sets of seven commuting operators are inequivalent. There is no problem to associate to each such operator set a bit chain. Only the bit chain associated to the set $\hat Q_1=(\tau_3\otimes\tau_1\otimes\tau_1)~,~\hat Q_2=(\tau_1\otimes\tau_3\otimes\tau_1)~,~\hat Q_3=(\tau_1\otimes\tau_1\otimes\tau_3)~,~\hat Q_2\hat Q_3~,~\hat Q_3\hat Q_1~,~\hat Q_1\hat Q_2~,~\hat Q_1\hat Q_2\hat Q_3=-(\tau_3\otimes\tau_3\otimes\tau_3)$ should be a new bit chain with quantum observables $(Q_1,Q_2,Q_3~,~Q_2\cdot Q_3~,~Q_3\cdot Q_1~,~Q_1\cdot Q_2~,~Q_1\cdot Q_2\cdot Q_3)$ which are different from the candidate $Q$-chain $({\stackrel \approx F}, {\stackrel \approx G},{\stackrel \approx H},\tilde Q_1,\tilde Q_2,\tilde Q_3,{\stackrel \approx Q})$ discussed above. This does not lead to any contradiction, since the map from quantum observables to operators is not invertible. Both the observables $Q_1$ and ${\stackrel \approx F}$ are mapped to the same operator $\hat Q_1$, but the classical product may be a quantum observable for $Q_1\cdot Q_2$ and not for ${\stackrel \approx F}\cdot Q_2$. In our explicit construction of a classical representation of observables we should not exclude $\hat Q_1$ from the set of independent operators with the argument that it can be obtained as the product of two commuting observables.

\medskip
\noindent
{\bf Classical entanglement}

Entanglement is a key feature of quantum mechanics. Its classical realization is best discussed in the context of bit chains. Consider $M=4$ and a bit chain of observables $T_1,T_2,T_3$, with corresponding commuting diagonal operators $L_1,L_2,L_3=L_1L_2$. We associate a first bit to $T_1$ and a second bit to $T_2$. The product of the two bits is then determined by the classical product $T_1\cdot T_2=T_3$. Let us concentrate on a state with $\rho_3=\kl T_3\kr=-1$ , $\rho_1=\kl T_1\kr=0~,~\rho_2=\kl T_2\kr=0$. Depending on the other $\rho_k$ this may be a pure or mixed state, with purity $P=1+\sum_{k\geq 4}\rho^2_k$. From $\rho_1=0$ we infer an equal probability to find for the first bit the values $+1$ and $-1$, and similar for the second bit from $\rho_2=0$. On the other hand, $\rho_3=-1$ implies a maximal anticorrelation  between bits one and two. The probabilities vanish for all classical states for which both bit one and bit two have the same value corresponding to $w_{++}=w_{--}=0$~,~$w_{-+}=w_{-+}=1/2$. 

We may assume that a first apparatus measures bit one, and a second one bit two. Whenever the first apparatus shows a positive result, the second apparatus will necessarily indicate a negative result, and vice versa. By itself, this anticorrelation does not yet indicate an entangled pure state. For example, it may be realized by a mixed state with $\rho_k=0$ for $k\geq 4$ , $P=1$, corresponding to a diagonal density matrix $\rho=(1/2)diag(0,1,1,0)$. 

We may compute the quantum correlation between a rotated spin observable $A(\vartheta)$ with associated operator $\hat A(\vartheta)=\cos\vartheta L_1+\sin\vartheta L_8$ and a second rotated spin observable $B(\varphi)$ with $\hat B(\varphi)=\cos\varphi L_2+\sin \varphi L_4$. One finds for arbitrary $\rho_k$ 
\ba\label{37}
\kl A(\vartheta)B(\varphi)\kr&=&C(\vartheta,\varphi)=\frac12\tr \big(\{\hat A(\vartheta),\hat B(\varphi)\}\rho\big)\nn\\
&=&\cos\vartheta\cos\varphi\rho_3+\cos\vartheta\sin\varphi\rho_6\\
&&+\sin\vartheta\cos\varphi\rho_{10}+\sin\vartheta\sin\varphi\rho_{12},\nn
\ea
where we use the representations
\ba\label{38}
L_1&=&(\tau_3\otimes 1)~,~L_2=(1\otimes\tau_3)~,~L_3=(\tau_3\otimes\tau_3),\nn\\
L_8&=&(\tau_1\otimes 1)~,~L_4=(1\otimes\tau_1)~,~L_{12}=(\tau_1\otimes\tau_1),\\
L_6&=&(\tau_3\otimes\tau_1)~,~L_{10}=(\tau_1\otimes\tau_3)~,~L_{14}
=-(\tau_2\otimes\tau_2).\nn
\ea
For all states with $\rho_3=\rho_{12}=-1~,~\rho_6=\rho_{10}=0$, one obtains the familiar quantum result for two spins with relative rotation
\be\label{39}
\kl A(\vartheta)B(\varphi)\kr=-\cos(\vartheta-\varphi)=\bar C(\vartheta-\varphi).
\ee
Bell's inequality for local deterministic theories reads for this situation
\be\label{40}
|C(\vartheta_1,0)-C(\vartheta_2,0)|\leq 1+C(\vartheta_1,\vartheta_2).
\ee
With eq. \eqref{39} this reduces to $|\bar C(\vartheta_1)-\bar C(\vartheta_2)|\leq 1+\bar C(\vartheta_1-\vartheta_2)$. It is violated for $\vartheta_1=\pi/2~,~\vartheta_2=\pi/4$. 

We observe that the contribution $\sim\rho_{12}$ to the quantum correlation matters. For our choice $\vartheta_1=\pi/2~,~\vartheta_2=\pi/4$ the inequality \eqref{40} reads 
$|C(\pi/2,0)-C(\pi/4,0)|\leq 1+C(\pi/2,\pi/4)$. For $\rho_6=\rho_8=0~,~\rho_3=-1$ and general $\rho_{12}$ one finds $C(\pi/2,0)=0~,~C(\pi/4,0)=-1/\sqrt{2}$ and $C(\pi/2,\pi/4)=\rho_{12}/\sqrt{2}$. For $\rho_{12}=0$ Bell's inequality is now obeyed.

We conclude that the presence of off-diagonal elements in the density matrix (in a direct product basis for the two entangled spins) plays an important role for the coexistence of different complete bit chains. For our example with $M=4$ a second bit chain besides $T_1,T_2,T_3=T_1\cdot T_2$ is given by $T_8,T_4,T_{12}=T_8\cdot T_4$, with associated commuting operators $L_8,L_4,L_{12}=L_8L_4$. Not only the spins in one direction are maximally anticorrelated for $\rho_3=-1$, but also the spins in an orthogonal direction (represented by $L_8,L_4)$ are maximally anticorrelated for $\rho_{12}=-1$. The quantum state of the subsystem allows for a specification of several correlations by independent elements as $\rho_3$ and $\rho_{12}$. This possibility is closely connected to the use of quantum correlations for the calculation of the outcome of two measurements. In a setting where only the classical correlations are available, a simultaneous implementation of the two two-bit chains $(T_1,T_2,T_3)$ and $(T_8,T_4,T_{12})$ would require more than three mutually commuting objects and can therefore not be implemented for a state with purity $P\leq 3$.

We finally display the classical formulation for two particular entangled pure states. They are given by 
\be\label{41}
\rho_3=\epsilon\rho_{12}=-\epsilon\rho_{14}=-1~,~\epsilon=\pm 1.
\ee
The sign $\epsilon=+1$ corresponds to the rotation invariant spin singlet state with density matrix
\be\label{42}
\rho=\frac14\big(1-(\tau_1\otimes\tau_1)-(\tau_2\otimes\tau_2)-(\tau_3\otimes\tau_3)\big),
\ee
and wave function
\be\label{43}
\psi=\frac{1}{\sqrt{2}}(\hat\psi_2-\hat\psi_3).
\ee
For $\epsilon=-1$ the relative sign between $\hat\psi_2$ and $\hat\psi_3$ is positive. 

\medskip
\noindent
{\bf Conclusions}

We have obtained all laws of quantum mechanics from classical statistics, including the concept of probability amplitudes $\psi$ and the associated superposition of states with interference and entanglement, as well as the unitary time evolution. Our classical statistical description is genuinely probabilistic and not a local deterministic model. It allows to predict probabilities for the outcome of a chain of measurements, but not a deterministic result of a given measurement in terms of some ``hidden variables''. Bell's inequalities do not apply for our formulation of a correlation function which is based on conditional probabilities for a sequence of measurements. Since the mapping from classical observables to quantum operators is not invertible, no contradiction to the Kochen-Specker theorem arises.

Our setting can be extended to include observables like location and momentum by considering many two-level observables on a space-lattice and taking the limit of vanishing lattice spacing. In our statistical mechanics setting quantum mechanics describes isolated subsystems of a larger ensemble that also includes the environment. Isolation does not mean that the subsystem can be described by classical probabilities for the states of the subsystem and sharp values of the observables in these states. It rather relates to a separated time evolution of the subsystem and to observables which can be described by quantities only associated to the subsystem, without explicit reference to the environment. 

We do not intend to enter here the debate if quantum mechanics or classical statistics are more fundamental -it is well known that classical statistics can be obtained as a limiting case of quantum mechanics. In our view classical statistics and quantum mechanics are two sides of the same medal. This may have far reaching consequences, as the possibility that the late time asymptotic state of a classical ensemble is given by the equilibrium ensemble of quantum statistics, or that certain steps in quantum computations can find a classical analogue. We find it remarkable that the conceptual foundations of quantum mechanics need not to go beyond the concepts of classical statistics.


\end{document}